\def\be{\begin{equation}}
\def\ee{\end{equation}}
\def\kt{k_{\perp}}
\def\pt{p_{\perp}}

\documentclass[12pt,a4paper]{article}
\usepackage{graphicx}
\usepackage{soul}

\usepackage[intlimits]{amsmath}
\usepackage{slashed}
\usepackage{graphicx}
\usepackage{subfigure}
\usepackage[svgnames]{xcolor}
\usepackage[normalem]{ulem}

\usepackage{amsmath,amssymb,amsfonts,graphicx}
\usepackage{bm}
\usepackage{slashed}
\usepackage[caption=false]{subfig}
\usepackage[svgnames]{xcolor}

\newcommand{\bea}{\begin{eqnarray}}
\newcommand{\eea}{\end{eqnarray}}

\begin{document}
\textwidth=135mm
 \textheight=200mm
\begin{center}
{\bfseries Extracting TMDs from CLAS12 data}

%\footnote{{\small Talk at the SPIN2012, JINR, Dubna, July 7 - 9,2005.}}
\vskip 5mm
M. Aghasyan$^{\dag}$ and H. Avakian$^\ddag$
\vskip 5mm
{\small {\it $^\dag$ LNF-INFN, Via E. Fermi 40, Frascati, Italy}}
\\
{\small {\it $^\ddag$ JLab, 12000 Jefferson Ave, Newport News, VA 23606, USA}} \\

\end{center}
\vskip 5mm
\centerline{\bf Abstract}
We present studies of double longitudinal 
spin asymmetries in semi-inclusive deep inelastic 
scattering  using a new dedicated Monte Carlo generator, which includes quark intrinsic transverse momentum
 within the generalized parton model 
based on the fully differential cross section for the process. 
Additionally we employ Bessel-weighting to the 
MC events to extract transverse momentum dependent 
parton distribution functions 
and also discuss possible uncertainties due to kinematic correlation effects.

\vskip 10mm
\section{\label{sec:intro}Fully differential SIDIS cross section}
The study of the 3-dimensional structure of protons and neutrons is one of the central issues in hadron physics, with many dedicated experiments, either running (COMPASS at CERN, CLAS and Hall-A at JLab, STAR and PHENIX at RHIC), approved (JLab $12~{\rm GeV}$ upgrade, COMPASS-II) or being planned (ENC/EIC Colliders). The transverse momentum dependent (TMD) partonic distribution (PDF) and fragmentation functions (FF) play a crucial role in gathering and interpreting information towards a true 3-dimensional imaging of the nucleons. TMDs can be accessed in several experiments, but the main source of information is semi-inclusive deep inelastic scattering (SIDIS) of polarized leptons off polarized nucleon. 
For SIDIS,  the theoretical formalism is described in a series of papers~\cite{Anselmino:2005nn,Boglione:2011wm} using tree level factorization~\cite{Mulders:1995dh} where  the standard momentum convolution integral~\cite{Bacchetta:2006tn} relates the quark
intrinsic transverse momentum to the  transverse momentum of the produced hadron $P_{hT}$ in semi-inclusive processes.
%integration over $k_{\perp}$ and $p_{\perp}$ using 
%famous $\delta^2(k_{\perp}^2z^2+p_{\perp}^2-P_{h,T}^2)$ function, which gives rise of the above mentioned convolution. 

In this work we present a model independent extraction of the ratio of polarized, $g_1$, and umpolarized, $f_1$, TMDs using
a Monte Carlo (MC) based on fully differential 
cross section, in which we re-construct the  final hadron transverse momentum 
after MC integration over the intrinsic quark  transverse momenta. 
In the MC generator we used the model described in Ref. \cite{Anselmino:2005nn} that was  numerically further evolved 
in  Ref. \cite{Boglione:2011wm}. 
The Bessel-weighted asymmetry, providing access to the ratio of Fourier transforms of $g_1$ and  $f_1$, 
has been extracted. The uncertainty of the extracted TMDs was estimated using different input models for distribution and fragmentation functions.

%\section{Fully differential SIDIS Cross section}

A fully differential Monte-Carlo generator has been developed to describe the the  SIDIS process when a final state hadron is detected with the final state lepton,
\be
{\ell}(l) + N(P)\rightarrow \ell(l') + h(P_{h}) + X, 
\ee
where $\ell$ is the lepton, $N$ the proton target and $h$ the observed hadron (four-momenta notations are given in parentheses). The virtual photon momentum is defined $q = l-l'$ and its virtuality $Q^2=-q^2$.

The fully differential SIDIS cross section used in MC is given by \cite{Anselmino:2005nn}:

\bea
\frac{d \sigma}{ dx dy dz d {\bf p^2_{\perp}}  d {\bf k^2_{\perp}}} &=& K\, \left[  \sum_q J \left \{ f_{q}(x,k_{\perp})D_{q,h}(z,p_{\perp})  + \lambda \sqrt{1-\epsilon}g_{q}(x,k_{\perp})D_{q,h}(z,p_{\perp}) \right \} \right], 
\label{FDXS}
\eea
where the summation runs over the quark flavors and $\epsilon$, $K(x,y)$ and  $J(x,Q^2,\kt)$ are some kinematic factors defined by the elementary scattering process \cite{Anselmino:2005nn},
$x$ is the Bjorken variable,  $k_{\perp}$ is the initial quark transverse 
momentm, $p_{\perp}$ is the transverse momentum of the final hadron with respect to scattered quark, and $y$ and $z$ are the fractional energies of the virtual photon and  detected hadron. 
For our studies we used simple factorized Gaussians for the $f_1(x,k_{\perp})$ and  $g_1(x,k_{\perp})$ distribution functions and  $D_1$ fragmentation function,
with widths given by fits from available world data.

\begin{figure}[t]
\centerline{\includegraphics[height=2.5in]{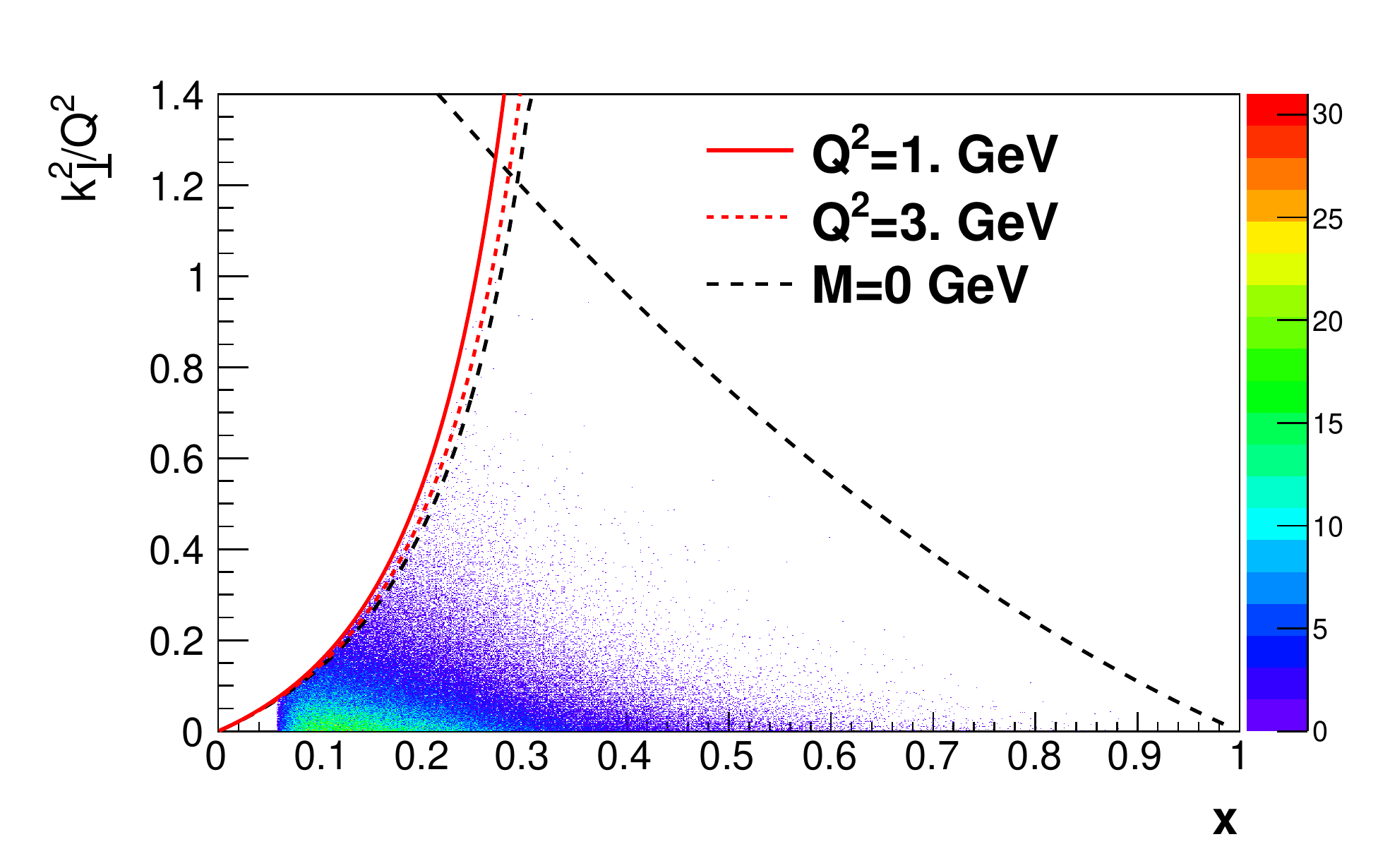}}
\caption{\scriptsize The $k^2_{\perp}/Q^2$ versus $x=x_B$ is presented for $11 \,{\rm GeV}$ electron beam. Black dashed lines are from \cite{Boglione:2011wm}, with proton mass zero approximation. Events above black dashed curve are due to the hadron mass. Red curves are calculated under the $k_z>0$ condition with non-zero hadron mass and for two $Q^2$ values.  Electron beam energy is $11 \,{\rm GeV}$.}
\label{ktQ2vsxb11}
\end{figure}

Figure~\ref{ktQ2vsxb11} shows the two dimensional 
plot of $k^2_{\perp}/Q^2$ versus $x$. 
One can see a clear correlation between the transverse and longitudinal momenta, 
where the dashed black curves \cite{Boglione:2011wm} defining the upper bounds. Restrictions at large $x$ come from
the  energy and momentum conservation and at small $x$ from the requirement that the parton should move in the 
forward direction with respect  to the  parent hadron ($k_z>0)$. 
The red curves (solid for $Q^2=1~{\rm GeV}^2$ and dashed $Q^2=3~{\rm GeV}^2$) 
are  calculated for a non-zero proton mass, requiring the parton to move in the forward direction with 
respect  to the parent hadron (in Ref.~\cite{Boglione:2011wm}  the proton mass was neglected). 

We note that  the portion of the data above the black dashed curve 
decreases  with decreasing $x$. Thus, the smaller the $x$, 
the smaller is the difference between red solid and red dashed curves 
implying  a diminishing role  of the hadron mass at higher  $Q^2$.
 
\section {Bessel-weighted extraction of the double spin asymmetry $A_{LL}$}

Extraction of the double spin asymmetry $A_{LL}$, defined as the ratio of difference and sum of electroproduction cross sections for antiparallel, $\sigma^+$, and parallel, $\sigma^-$, configurations of lepton and nucleon spins, has been performed using the Bessel-weighting procedure described in Ref. \cite{Boer:2011xd}.
Within that approach one can extract the Fourier transform of the double spin asymmetry, $ A^{J_{0}(b_TP_{hT})}_{LL}(b_T)$, using measured double spin asymmetries as a function of the $P_{hT}$~\cite{Avakian:2010ae},  for fixed $x$, $y$, and $z$ bins.
\be
 A^{J_{0}(b_TP_{hT})}_{LL}(b_T) = \frac{ \tilde \sigma^+(b_T) - \tilde \sigma^-(b_T)}{\tilde \sigma^+(b_T) + \tilde \sigma^-(b_T) }=\frac{\tilde \sigma_{LL}(b_T)}{\tilde \sigma_{UU}(b_T)}=\sqrt{1-\varepsilon^2} \frac{\sum_{q}\tilde g^{q}_1(x,z^2b_T^2) \tilde D^{q}_{1}(z,b_T^2)}{\sum_{q}\tilde f^{q}_1(x,z^2b_T^2) \tilde D^{q}_{1}(z,b_T^2)},   
 \label{tildall}
\ee 
where $b_T$ is the Fourier conjugate of the $P_{hT}$. The Fourier transforms of helicity dependent cross sections, $\sigma^{\pm}( b_T)$, can be extracted by integration (analytic models) or summation (for data and MC) over the hadronic transverse momentum, weighted 
by a Bessel function $J_{0}$,
\be
 \tilde \sigma^{\pm}( b_T) \simeq S^{\pm}=\sum_{i=1}^{N^{\pm}} J_{0}(b_T P_{hT,i}) \,  .
 \label{spm}
\ee

\begin{figure}[t]
\centerline{\includegraphics[height=2.5in]{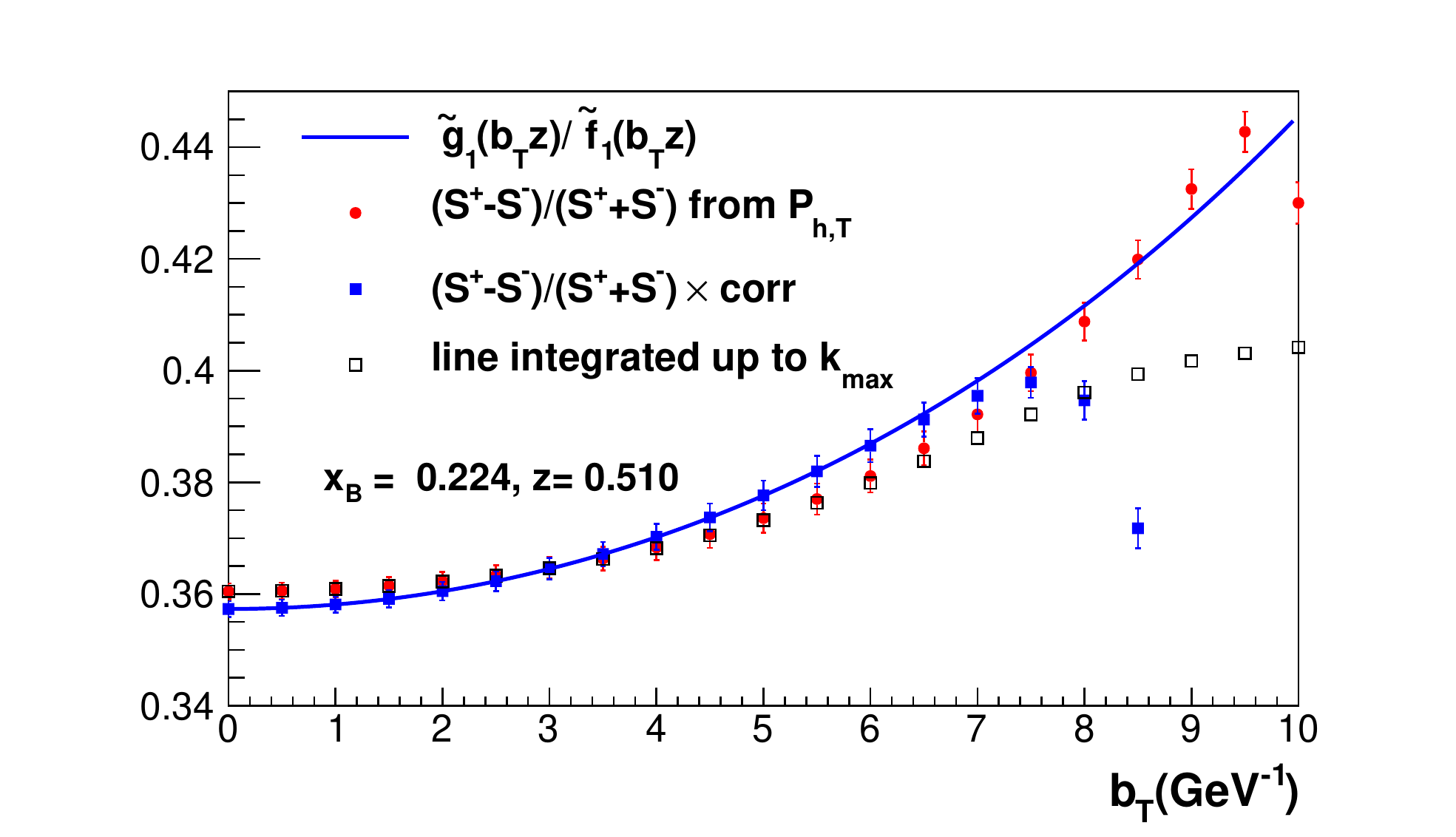}}
\caption{\scriptsize Extracted Bessel-weighted asymmetry versus $b_T$ with and w/o correction, compared to values 
calculated analytically and numerically directly from the input. The MC sample was produced assuming simple Gaussian 
DFs and FFs with  $<k^2_{\perp}>_{g_1} = 0.8<k^2_{\perp}>_{f_1}$ and $g_1(x) = f_1(x)x^{0.7}$.
}
\label{BWPhTcorrg1f1}
\end{figure}

In the Fig.~\ref{BWPhTcorrg1f1} the red points represent the outcome of  the 
Bessel-weighted asymmetries from the  MC sample, while the blue curve represents the analytical 
expression $\frac{\tilde g_1(x,zb_T)}{\tilde f_1(x,zb_T)}$ using $<\kt^2>_{g_1}$ and $<\kt^2>_{f_1}$ 
from the fits to $\kt^2$ distributions from the same MC sample.

Within the $b_T$ range of $b_T<5\sim 6~{\rm GeV}^{-1} \simeq 1~fm$  the  Bessel-weighted asymmetries could be extracted with a minimum of  2.5\% accuracy, although with some systematic shift. 
The shift observed in the reconstructed value  is due to the  kinematical restriction introduced by energy and momentum conservation~\cite{Boglione:2011wm}, which deforms  the Gaussian shapes 
of the $\kt$ and $\pt$ distributions.
 In experiment there is always a cutoff at high $P_{hT}$ due to the acceptance and low cross section,  as well as a cutoff at small $P_{hT}$, where the azimuthal angles are not well defined. 
A correction factor, accounting for the missing
$P_{hT}$ range above the maximum value accessible ($P_{max}$) in a given experiment, was estimated,  based on an analytic calculation of the contribution above that value.
The systematic uncertainty of that correction could be estimated from the variation of the maximum $P_{max}$ within the resolution of the experiment. 
In Fig.~\ref{BWPhTcorrg1f1} the blue filled squares represent this correction using the Gaussian distributions and the open black squares represent the 
numerical integration up to that exact 
$k^2_{max}$ that we used for correction.

This work was supported by the Research Infrastructure Integrating Activity
Study of Strongly Interacting Matter (acronym HadronPhysic3, Grant Agreement
n. 283286) under the Seventh Framework Programme of the European Community.

%===============================================================================
%===============================================================================
 %===============================================================================
%===============================================================================

\end{document}